\documentclass[12pt]{article}
\usepackage[margin=2cm]{geometry}
\usepackage{comment}
\usepackage{amsmath,amssymb,extarrows,mathtools,graphicx,subfigure,setspace}
\usepackage{cite}
\usepackage{slashed}
\usepackage{color}
\usepackage{tikz}
\usepackage{fancyhdr}
\usetikzlibrary{decorations.pathmorphing}
\makeatother

\newcommand{\be}{\begin{equation}}
\newcommand{\bea}{\begin{eqnarray}}
\newcommand{\eea}{\end{eqnarray}}
\newcommand{\ba}{\begin{array}}
\newcommand{\ea}{\end{array}}
\newcommand{\ee}{\end{equation}}
\newcommand{\bes}{\begin{equation*}}
\newcommand{\beas}{\begin{eqnarray*}}
\newcommand{\eeas}{\end{eqnarray*}}
\newcommand{\bas}{\begin{array*}}
\newcommand{\eas}{\end{array*}}
\newcommand{\ees}{\end{equation*}}

\numberwithin{equation}{section}
\begin{document}
	\onehalfspacing
	\noindent

\begin{titlepage}
\vspace{10mm}
\begin{flushright}

\end{flushright}

\vspace*{20mm}
\begin{center}

{\Large {\bf  Complexity  and  Near Extremal Charged Black Branes}\\
}

\vspace*{15mm}
\vspace*{1mm}
{Mohsen Alishahiha${}^\ast$,  Komeil Babaei Velni$^{\dagger}$ and 
 Mohammad Reza Tanhayi${}^{\ddagger, \ast}$}

 \vspace*{1cm}

{\it ${}^\ast$ School of Physics,
Institute for Research in Fundamental Sciences (IPM)\\
P.O. Box 19395-5531, Tehran, Iran\\
 ${}^\dagger$ Department of Physics, University of Guilan,
P.O. Box 41335-1914, Rasht, Iran\\
${}^\ddagger$ Department of physics, Islamic Azad University  Central Tehran Branch, Tehran, Iran\\
}

 \vspace*{0.5cm}
{E-mails: {\tt alishah@ipm.ir, babaeivelni@guilan.ac.ir, mtanhayi@ipm.ir}}%

\vspace*{1cm}
\end{center}

\begin{abstract}
We compute holographic complexity of charged black brane solutions in 
arbitrary dimensions for the  near horizon limit  of near extremal case using  
two different methods. The corresponding complexity may be 
obtained either by taking the limit from the complexity of the charged black brane, or by
computing the complexity for near horizon limit of near extremal solution. One observes that these
results coincide if one assumes to have a cutoff behind 
horizon whose value is fixed by UV cutoff and also taking into account a proper  counterterm
evaluated on this cutoff. We also consider the situation for  Vaidya charged black branes too.

\end{abstract}

\end{titlepage}

\section{Introduction}

In the context of black hole physics, horizon, black hole entropy, information paradox and  
physics behind the horizon are in the center of the most studies in last  several decades. 
Quantum information theory might be capable to shed light on these subjects. Indeed,
recent progress on black hole physics has opened up a possibility to make a connection 
between quantum information theory and back hole physics (see \cite{Almheiri:2012rt}
and its citations).  To explore and understand  this possible connection the AdS/CFT
correspondence \cite{Maldacena:1997re} has played rather an important role. 
In this context holographic entanglement entropy \cite{Ryu:2006bv} and computational 
complexity\cite{Susskind:2014rva,Stanford:2014jda} may be  thought of as  
examples which could make this connection more concrete.   

In particular, holographic complexity, by definition, might be able to give us some 
information on the physics behind the horizon. To elaborate this point better it is 
worth noting that  the holographic complexity  may be obtained 
by the on shell action evaluated on a certain subregion of spacetime known as 
the Wheeler-DeWitt (WDW) patch\cite{{Brown:2015bva},{Brown:2015lvg}}, that 
includes some portion of the spacetime located behind  the horizon. 

More precisely, in this picture known as ``complexity equals to action'' (CA)  proposal 
 \cite{{Brown:2015bva},{Brown:2015lvg}} the late time behavior of complexity growth is 
 entirely given by the on shell action evaluated on the intersection  of the WDW patch with 
 the future interior\cite{Alishahiha:2018lfv}, leading  to an observation that the late time 
 behavior of holographic complexity  is insensitive to the UV cutoff \cite{Akhavan:2018wla}.  
 This, in turn, results in a conclusion  
that there could be a relation between the UV cutoff and a cutoff that should be  defined 
behind the horizon \cite{{Akhavan:2018wla}, {Alishahiha:2018swh}} whose the value is 
fixed by the UV cutoff. Indeed this is an interesting feature  of complexity that could probe 
physics behind the horizon.

It is important to mention that the result of \cite{Akhavan:2018wla}
leading to the behind the horizon cutoff  relies on two facts. The first one is what we 
just mentioned, namely,   the late time behavior of complexity   is UV blind. The 
second one is that, according to Lloyd's bound  \cite{Lloyd:2000}, the late time behavior
of the complexity growth  is given in terms of  the energy of the system that is sensitive to 
the finite UV cutoff. 

We note, however, that  in the context of holographic complexity it is known that 
the Lloyd's bound may be  violated \cite{{Carmi:2017jqz},{Kim:2017qrq},
{Couch:2017yil},{Swingle:2017zcd},
{An:2018xhv},{Alishahiha:2018tep}}. Nonetheless, the violation
of Lloyd's bound  just modifies the relation between the cutoff behind the horizon and 
the  UV cutoff and the Lloyd's bound violation does not affect the main conclusion. 
 Indeed, the only fact we need  is that the late time behavior of complexity growth is 
 controlled  by physical charges  defined at the boundary of which the values are affected 
 by a finite UV cutoff.

This is the aim of the present paper to further  explore the significance of the cutoff  behind 
the horizon. To do so, we will study holographic complexity for AdS-Reissner-Nordstr\"om
(AdS-RN) black branes in arbitrary dimensions (see also 
 \cite{{Cai:2016xho},{Carmi:2017jqz},{Ovgun:2018jbm},{Ghaffarnejad:2018prc}}). More
 precisely, we will consider near horizon limit of near extremal  charged black branes 
 and compute  complexity in two different ways. In the first approach we will
 obtain the corresponding  rate of complexity growth by taking the near extremal
  limit of the complexity growth of the charged black brane. In the second 
  approach  we compute complexity for a metric obtained by taking 
 near horizon  limit of the near extremal  AdS-RN black brane that has the
 form of AdS$_2\times \mathbb{R}^{d-1}$.

It is important to note that  if one naively computes complexity for geometries containing 
an AdS$_2$ factor the late time growth vanishes
 \cite{{Akhavan:2018wla},{Brown:2018bms},{Goto:2019}}
 that is  not consistent with what we may get from taking near extremal limit of the
 complexity growth.  On the other hand, as we will see,  following  \cite{Akhavan:2018wla} if one 
considers the cutoff behind the horizon whose value  is fixed by the UV cutoff, then
one finds the standard late time linear growth that is exactly the one obtained by the first 
approach. It is, however, important to note that in order to get a consistent result one needs also
to consider all counter terms already obtained in the context of holographic renormalization
\cite{Alishahiha:2018swh}. 

Indeed, we believe  that the result of the present paper  should be considered as an 
evidence supporting the proposed idea  that the UV cutoff will  automatically fix a 
cutoff behind the horizon. Moreover, it also explores the role of the counter terms 
which are  required by  holographic renormalization.
 
The organization of the paper is as follows. In the next section in order to fix our notations
we will review the computation of  complexity for charged black branes. In section three
we will study  near horizon limit of  near extremal cases where we compute the 
corresponding complexity from 
two different approaches. In section four we will redo the same computations for charged Vaidya 
black brane. The last section is devoted to summery and conclusions.


\section{Complexity of charged black branes: a brief review}

In this section in order to fix our notation we will review holographic complexity for 
charged black branes using CA proposal. Actually the full time 
dependence of the complexity growth for charged black holes has been first 
studied in \cite{Carmi:2017jqz}. The aim of this section is, while  reviewing 
the CA proposal for charged black branes,  presenting  a closed inspiring  
form for the corresponding  holographic complexity. 
To do so, we will use an appropriate coordinate system 
that simplifies the computations  rather significantly.  Although the result is known the presentation 
is new and by itself is interesting.

To proceed,  let us consider an Einstein-Maxwell theory in $d+1$ dimensions of which
the action may be given as follows 
\be\label{action0}
I_0=\frac{1}{16\pi G_N}\int d^{d+1}x\sqrt{-g}\left(R+\frac{d(d-1)}{L^2}
-\frac{1}{2}F_{\mu\nu}F^{\mu\nu}\right)\,,
\ee
where $G_N$ is the Newton's constant and $L$ is a length scale in the theory.
The corresponding equations of motion are
\bea\label{EOM}
R_{\mu\nu}-F_{\alpha\mu}F^{\alpha}_{\,\nu}-\frac{1}{2}\left(R+\frac{d(d-1)}{L^2}-\frac{1}{2}F^2\right)g_{\mu\nu}=0,\;\;\;\;\;\;\;\partial_\alpha\left(\sqrt{-g}F^{\alpha\beta}\right)=0\,.
\eea
These equations  admit  AdS-RN  black brane solutions  that for $d\geq 3$ are given by  
\bea\label{SOL}
&&ds^2=\frac{L^2}{r^2}\left(-f(r)dt^2+\frac{dr^2}{f(r)}+\sum_{i=1}^{d-1}d{x}_i^2\right),\;\;\;\;\;
f(r)=1-m r^d+{Q^2} r^{2(d-1)},
\cr &&\cr
&&A_t=\sqrt{\frac{d-1}{d-2}}QL \;(r_+^{d-2}-r^{d-2}),
\eea
where $m$ and $Q$ are related to the mass and charge of the black brane 
solution, respectively.  Moreover, $r_+$ is the radius of horizon  that is the smallest 
solution of  $f(r)=0$ ( note that in our notation the boundary is at
$r=0$). For $d=2$ the solution  exhibits a logarithmic behavior.

Now, the aim is to compute   holographic complexity for the above charged black brane 
using  CA proposal. To do so,  one needs to compute on shell action on 
the WDW patch associated with 
a boundary state given at $\tau=t_L+t_R$. Here $t_L (t_R)$, is  time coordinate of left (right)
boundary of the charged black brane (see left panel of the figure 1).  To proceed, we note that the action consists of several parts  including bulk, boundary and joint 
points as follows \cite{{Parattu:2015gga},{Parattu:2016trq},{Lehner:2016vdi}}
\bea\label{ACT0}
I&=&I_0+\frac{1}{8\pi G_N}
\int_{\Sigma^{d}_t} K_t\; d\Sigma_t \pm\frac{1}{8\pi G_N} \int_{\Sigma^{d}_s} K_s\; d\Sigma_s\pm \frac{1}{8\pi G_N} 
\int_{\Sigma^{d}_n} K_n\; dS 
d\lambda\cr &&\cr &&\pm\frac{1}{8\pi G_N} \int_{J^{d-1}} a\; dS\pm
\frac{1}{8\pi G_N}\int_{\Sigma^{d}_n} d\lambda dS \,\Theta\log\frac{|L\Theta|}
{d-1}
\,.
\eea
Here the timelike, spacelike, and null boundaries and also joint points are denoted by $
\Sigma_t^{d}, \Sigma_s^{d}, \Sigma_n^{d}$ and $J^{d-1}$, respectively. The extrinsic 
curvature of the corresponding boundaries are given by $K_t, K_s$ and $K_n$. The function $a$ 
at the intersection of the boundaries is given by the logarithm of the inner product of the 
corresponding normal vectors and $\lambda$ is the null coordinate defined on 
the null segments. The sign of different terms depends on the relative position 
of the boundaries and the bulk region of interest (see \cite{Lehner:2016vdi} for more details).
Moreover, in the last term the quantity $\Theta$ is defined as follows
\be
\Theta=\frac{1}{\sqrt{\gamma}}\frac{\partial\sqrt{\gamma}}{\partial\lambda},
\ee
where $\gamma$ is determinant of induced metric on the joint points.  This term is needed 
in order  to remove the ambiguity associated with the normalization of   null vectors
\cite{{Lehner:2016vdi},{Reynolds:2016rvl}}.  As we will see this term together with 
other counterterms play crucial role in order to get desired results.

 It is worth noting that in general to write the last term one could 
use an arbitrary length scale, though for simplicity we have fixed  the corresponding scale 
 to be  the  radius of curvature $L$. This choice just removes the most divergent term 
from the  complexity and does not affect the physics that we are interested in.

It is also important to note that besides the above terms, there are also
other boundary terms which could contribute to the complexity. These terms are 
those counterterms needed to make the on shell action finite when evaluated over whole 
spacetime\cite{Emparan:1999pm}. 
The importance of these terms have been also known from holographic renormalization
point of view (see {\it e.g.}\cite{Bianchi:2001kw}). 
In the present case since the solution has flat boundary the
only term remains to be considered is (for asymptotically AdS$_{d+1}$  metric) 
\be\label{CT1}
I^{\rm c.t.}=\frac{1}{8\pi G_N}\int d\Sigma \,\,\frac{d-1}{L}\,.
\ee

Note that this term gives non-zero contributions to the holographic complexity whenever the
corresponding   WDW patch contains  space like or time like boundaries. 
More precisely,  for a time like boundary at
the UV region of the space time, typically,  this counter term needed to remove certain 
divergent  term. Of course it may also have a finite contribution to the complexity  
as well, though its  contribution does not change  the complexity growth. 
On the other hand, for a space like boundary (that typically appears behind the horizon) 
this term might lead to a time dependent contribution to the complexity. This is, indeed,
the point we are going to explore in the next section.

It is obvious that  for the  charged black brane \eqref{SOL} and the corresponding  WDW 
patch shown in the left panel of the figure 1, since all boundaries are null, this term vanishes identically, though as we will see for 
the case in which  the WDW patch contains 
a space like boundary this term gives a non-zero contribution to the on shell action (see eq.
\eqref{CT}).

\begin{figure}
\begin{center}
\includegraphics[scale=0.99]{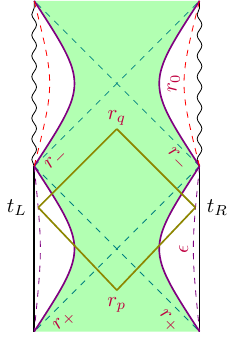}
\includegraphics[scale=0.99]{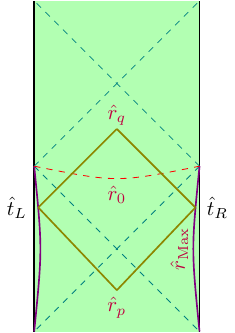}
\end{center}
\caption{{\it Left }: Penrose diagram of a charged black brane and the WDW 
patch associated with 
the sate defined at $\tau=t_R+t_L$. The green region shows the  Penrose diagram of  near 
extremal limit of the charged black brane. The red dashed lines is a timelike cutoff behaind
the horizon.  {\it Right}: Penrose diagram of an AdS$_2$ geometry 
which can be thought of as near horizon limit of the near extremal solution. The corresponding 
WDW patch is cut by the behind the horizon cutoff $\hat{r}_0$ that is fixed by the UV cutoff
$\hat{r}_{\rm Max}$.}
\label{fig:A}
\end{figure}

Let us now compute on shell action for the WDW patch depicted  
in the left panel of the figure 1. To do so, it is found useful to consider the following  
change of coordinate
\be
dt=dv+\frac{dr}{f(r)}\,,
\ee
by which the solution \eqref{SOL} may be recast into the following form
\be
ds^2=\frac{L^2}{r^2}\left(-f(r)dv^2-2 dr dv+d\vec{x}^2\right),\;\;\;\;\;\;
A_v=\sqrt{\frac{d-1}{d-2}}QL \;(r_+^{d-2}-r^{d-2})
\ee

In this notation the null boundaries of the corresponding WDW patch are given by 
$v=$constant  and $u=$constant. Here, the coordinate $u$ defined by $du=dt+\frac{dr}{f(r)}$.
Of course since we are using $(r,v)$ coordinate system one needs  to write $u$ coordinate
 in terms of $r$ and $ v$
 \be
du=dv+2\frac{dr}{f(r)}.
\ee
Therefore the null boundaries are given by $v=-t_L$ and $v=t_R$ for constant $v$ and,  for  
constant  $u$ one ends up with following equations for boundaries
\bea
&&\int_{-t_L}^v dv=-2\int_\epsilon^{r_q(v,\tau)} \frac{dr}{f(r)}\,\,\,\,
\rightarrow\,\,\,v+t_L=-2\int_\epsilon^{r_q(v,\tau)} \frac{dr}{f(r)},\cr &&\cr
&&\int_{t_R}^v dv=-2\int_\epsilon^{r_p(v,\tau)} \frac{dr}{f(r)}\,\,\,\,
\rightarrow\,\,\,t_R-v=2\int_\epsilon^{r_p(v,\tau)} \frac{dr}{f(r)}.
\eea

Before proceeding to compute the on shell action it is useful to fix our notation for the null
vectors of the null boundaries of the corresponding WDW patch. Indeed, as we have already  
mentioned the boundaries are given at $u$ and $v$ constant. Therefore the corresponding 
null vectors are given by 
\be\label{NV}
k_1=\alpha\partial_v,\;\;\;\;\;\;\;\;\;k_2=\beta\partial_u=\beta\left(\partial_v
+\frac{2}{f(r)}\partial_r\right), 
\ee
where $\alpha$ and $\beta$ are two free parameters appearing due to the ambiguity of
the normalization of null vectors.  

For the charged black brane which we are considering one has
\be
\sqrt{-g}\left(R+\frac{d(d-1)}{L^2}-\frac{1}{2}F^2\right)=-2d\frac{L^{d-1}}{r^{d+1}}
+2(d-2)L^{d-1}Q^2 r^{d-3}\,,
\ee
by which the bulk part of the on shell action reads
\bea
I^{\rm bulk}&=&\frac{L^{d-1}}{8\pi G_N}
\int d^{d-1}x\; \int_{-t_L}^{t_R} 
dv\int_{r_p(v,\tau)}^{r_q(v,\tau)}dr\, \left(\frac{-d}{r^{d+1}}+Q^2(d-2)r^{d-3}\right)\cr &&\cr
&=&\frac{V_{d-1}L^{d-1}}{8\pi G_N}
 \int_{-t_L}^{t_R} dv \left(\frac{f(r_q(v,\tau))}{r_q^{d}(v,\tau)}-
\frac{f(r_p(v,\tau))}{r_p^{d}(v,\tau)}\right).
\eea
On the other had using the fact that $\frac{dr_p(v,\tau)}{d\tau}=\frac{1}{2}f(r_p(v,\tau))$
and $\frac{dr_q(v,\tau)}{d\tau}=-\frac{1}{2}f(r_q(v,\tau))$
one can perform the above integration resulting in
\be
I^{\rm bulk}=\frac{V_{d-1}L^{d-1}}{8\pi G_N}\left(\frac{2}{(d-1)r_p^{d-1}(\tau)}
+\frac{2}{(d-1)r_q^{d-1}(\tau)}-
\frac{4}{(d-1)\epsilon^{d-1}}\right)\,.
\ee
Here we have used the fact  that in our notation one has
$r_p(\tau)=r_p(-t_L,\tau), r_q(\tau)=r_q(t_R,\tau)$
and $r_p(t_R,\tau)=r_q(-t_L,\tau)=\epsilon$.

Since all boundaries are null their contributions  vanish using Affine parameterization for null direction, while for the joint points using \eqref{NV} one gets 
\bea\label{J0}
I^{\rm joint }&=&\frac{V_{d-1}L^{d-1}}{8\pi G_N} \left(\frac{\log\frac{\alpha\beta r^2_p(\tau)}
{L^2|f(r_p)|}}{r_p^{d-1}(\tau)}+\frac{\log\frac{\alpha\beta r^2_q(\tau)}{L^2|f(r_q)|}}{r_q^{d-1}(\tau)}-
2\frac{\log\frac{\alpha\beta \epsilon^2}{L^2f(\epsilon)}}{\epsilon^{d-1}}
\right)\cr &&\cr
&=&\frac{V_{d-1}L^{d-1}}{8\pi G_N} \left(\frac{\log\frac{\alpha\beta r^2_p(\tau)}
{L^2}}{r_p^{d-1}(\tau)}+\frac{\log\frac{\alpha\beta r^2_q(\tau)}{L^2}}{r_q^{d-1}(\tau)}-
2\frac{\log\frac{\alpha\beta \epsilon^2}{L^2}}{\epsilon^{d-1}}
-\frac{\log {|f(r_p)|}}{r_p^{d-1}(\tau)}-\frac{\log{|f(r_q)|}}{r_q^{d-1}(\tau)}
\right).
\eea
The last contribution to the on shell action we need to consider is the contribution of the  
term required 
to remove the ambiguity associated 
with the normalization of null vectors. There are indeed four null boundaries
whose contributions  to the  on shell action are\footnote{To find this expression we use 
the fact that $\Theta=\pm \alpha \frac{(d-1)r}{L^2}, {\rm or}\, \pm \beta \frac{(d-1)r}{L^2}$, and
$\frac{\partial r}{\partial\lambda}=\pm \alpha\frac{r^2}{L^2},\,{\rm or}
=\pm \beta\frac{r^2}{L^2}$ depending on which null boundary is taken. }
\bea
I^{\rm amb}&=&\frac{V_{d-1}L^{d-1}}{8\pi G_N}\left(
-\frac{\log\frac{\alpha\beta r^2_p(\tau)}{L^2}}{r_p^{d-1}(\tau)}
-\frac{2}{(d-1)r_p^{d-1}(\tau)}+\frac{\log\frac{\alpha\beta \epsilon^2}{L^2}}{\epsilon^{d-1}}
+\frac{2}{(d-1)\epsilon^{d-1}}\right)\cr &&\cr
&&+\frac{V_{d-1}L^{d-1}}{8\pi G_N}\left(
-\frac{\log\frac{\alpha\beta r^2_q(\tau)}{L^2}}{r_q^{d-1}(\tau)}
-\frac{2}{(d-1)r_q^{d-1}(\tau)}+\frac{\log\frac{\alpha\beta \epsilon^2}{L^2}}{\epsilon^{d-1}}
+\frac{2}{(d-1)\epsilon^{d-1}}\right)\,.
\eea
Now putting all terms together one finds the following expression for the  total on shell action 
evaluated in the WDW patch
\be
I^{\rm total}=-\frac{V_{d-1}L^{d-1}}{8\pi G_N} \left(\frac{\log {|f(r_p(\tau))|}}{r_p^{d-1}(\tau)}+
\frac{\log{|f(r_q(\tau))|}}{r_q^{d-1}(\tau)}
\right)\,.
\ee
Interestingly enough, the final result is very simple and indeed consists of the 
contributions of  two joint points $p$ and $q$. It is also easy to compute its time 
derivative to find
\bea
\frac{dI^{\rm total}}{d\tau}&=&\frac{V_{d-1}L^{d-1}}{16\pi G_N}
 \bigg(\frac{ {f'(r_q(\tau))}}{r_q^{d-1}(\tau)}-
\frac{{f'(r_p(\tau))}}{r_p^{d-1}(\tau)}\\ &&\cr
&&
\;\;\;\;\;\;\;\;\;\;\;\;\;\;\;\;\;\;+\frac{(d-1)}{r_p^{d}(\tau)}f(r_p(\tau))\log {|f(r_p(\tau))|}-\frac{(d-1)}{r_q^{d}
(\tau)}f(r_q(\tau))
\log {|f(r_q(\tau))|}\bigg),\nonumber
\eea
which at the late time  reads
\be\label{R1}
\frac{dI^{\rm total}}{d{\tau}}=\frac{V_{d-1}L^{d-1}}{16\pi G_N}
 \bigg(\frac{ {f'(r_-)}}{r_-^{d-1}}-
\frac{{f'(r_+)}}{r_{+}^{d-1}}\bigg)
\,.
\ee
Using the explicit form of the function $f(r)$ one gets the same  expression know in the 
literature  \cite{{Cai:2016xho},{Carmi:2017jqz},{Ovgun:2018jbm},{Ghaffarnejad:2018prc}}.
Clearly in the extremal case where $r_-=r_+$ the complexity growth at the late time vanishes.
Of course this is not the case for the near extremal black brane as we will study in the following 
section. 


\section{Complexity of near horizon limit of near extremal solutions}

In this section we would like to study  complexity of near extremal charged  black branes.
When one is  considering near extremal case, it usually comes with taking near horizon limit
too. Therefore to be more precise in what follows we will be considering complexity for 
 near horizon limit of the near extremal black brane. The Penrose diagram of 
the near extremal solution is shown by green region  in the left panel of figure 1, while 
that of the near horizon limit is depicted in the right panel. Note that taking the 
near horizon limit of the near extremal black brane one gets a geometry with an AdS$_2$ 
factor parameterized  by the Rindler coordinates.

In this section we shall compute the corresponding complexity  for near 
horizon of the near extremal  black brane case in two different ways:  Either by 
taking the near horizon limit of the complexity growth of the charged black hole, or by 
computing the complexity growth of the near horizon limit of the near extremal black hole.
One would expect that the resultant complexity growths match. Of course to compare the 
results one needs to carefully evaluate the near horizon limits.  

To proceed, it is useful to adapt a notation that is more 
appropriate  for studying the  near extremal case. 
Indeed,  since the blacking factor $f(r)$ has two roots at $r=r_-$ and $r=r_+$,
one may write
\be
f(r)=\left(1-\frac{r}{r_-}\right)\left(1-\frac{r}{r_+}\right)h(r)\,,
\ee
where $h(r)$ is a function of $r$ defined via the above equation with the assumption
$h(r_\pm)\neq 0$. In this notation the late time behavior of the complexity of the charged 
black brane may be recast into the following form
\be\label{LTQ}
\frac{dI^{\rm total}}{d{\tau}}=\frac{V_{d-1}L^{d-1}}{4 G_N}\frac{r_--r_+}{4\pi r_-r_+}
 \bigg(\frac{ {h(r_-)}}{r_-^{d-1}}+
\frac{{h(r_+)}}{r_{+}^{d-1}}\bigg)=S_-T_-+S_+T_+
\,,
\ee
with 
\be
T_\pm=\frac{r_--r_+}{4\pi r_-r_+}\,h(r_\pm),\;\;\;\;\;\;\;\;\;S_\pm=
\frac{V_{d-1}L^{d-1}}{4 G_N}\frac{ {1}}{r_\pm^{d-1}}\,,
\ee
being  temperature and entropy one may associate to  inner and 
outer horizons.
 
To study near horizon of the
near extremal solution it is useful to  define new coordinates $(\hat{r},\hat{t})$  as follows
(see for example \cite{Sen:2008vm})
\be\label{SC}
r=\frac{r_+-r_-}{2}\,\frac{\hat{r}}{L}+\frac{r_-+r_+}{2},\;\;\;\;\;\;\;\;\;\;
t=\frac{2r_-r_+}{r_--r_+}\,\frac{\hat{t}}{L}\,.
\ee
The limit is then defined by $r_-\rightarrow r_+$ keeping the new 
coordinates fixed. Note that in this
limit  setting $r_-\approx r_+=r_e$ one has $f(r_e)=f'(r_e)=0$ and thus 
\be
r_e^{d-1}=\sqrt{\frac{d}{d-2}}\,\frac{1}{Q},\;\;\;\;\;\;\;\;\;\;
h(r_e)=\frac{r^2_e}{2}f''(r_e)=d(d-1)\,.
\ee
It is then easy to see that in this limit the solution \eqref{SOL} reduces to
\be\label{NES}
ds^2=-\left(\frac{\hat{r}^2}{L^2}-1\right)h(r_e)d\hat{t}^2+
\frac{d\hat{r}^2}{\left(\frac{\hat{r}^2}{L^2}-1\right)h(r_e)}+\frac{L^2}{r_e^2}
d\vec{\hat{x}}^2,\;\;\;\;
F_{\hat{r}\hat{t}}=\frac{\sqrt{d(d-1)}}{L}\,,
\ee
that is an AdS$_2\times \mathbb{R}^{d-1}$ geometry  whose 
Ricci scalar is $R=-\frac{2h(r_e)}{L^2}$. It is easy to check that this is also a solution 
of equations of motion \eqref{EOM}.

Now  let us  study the late time behavior of the complexity for near horizon limit of 
the  near extremal case. This may be done by taking the above limit from the 
late time behavior of 
complexity of the charged black brane given in the equation \eqref{R1}. To take the
 limit it is important to note that the boundary time, $\tau$, should be also  
rescaled properly that is the same as  that for coordinate $t$ given in the equation \eqref{SC}.  
Therefore  for $d>2$ one gets
\be\label{LNE0}
\frac{dI^{\rm total}}{d\hat{{\tau}}}=\frac{V_{d-1}L^{d-2}}{4\pi G_N}\frac{h(r_e)}{r_e^{d-1}}
=2 S_{\rm NE} T_{\rm NE}
\,,
\ee
that should be compared with \eqref{LTQ}. Note that in this expression 
\be
T_{\rm NE}=\frac{h(r_e)}{2\pi L},\;\;\;\;\;\;\;\;\;S_{\rm NE}=\frac{V_{d-1}}{4G_N}\,\frac{L^{d-1}}
{r_e^{d-1}} ,
\ee
are the Hawking temperature and entropy of the near horizon solution \eqref{NES}.

Alternatively one could compute the corresponding  complexity growth directly form the 
near horizon solution \eqref{NES}. Since the solution we are interested in 
has an AdS$_2$ factor, one may reduce the theory along the extra $(d-1)$ dimensions  into 
two dimensions in which  we typically obtain a two dimensional  gravity that might be 
thought of as  a generalized  Jackiw-Teitelboim gravity with non-zero background  charge.  
Of course we could still work within the  higher dimensional gravity and compute the 
complexity for a solution in the form of AdS$_2\times \mathbb{R}^{d-1}$. Indeed this is what we 
will do in what follows.
  
Before getting into details, it is worth recalling that by action we mean all terms required to 
have a general covariant action with well defined variational principle that results in a
finite free energy. In particular this means that, beside the 
Gibbons-Hawking term, one should consider all counter terms obtained in 
the context of holographic renormalization (see {\it e.g.} \cite{Skenderis:2002wp}). 
In particular in the present case the corresponding counter term is given by \eqref{CT1}.
As we will see this term plays an important role.
  
To proceed, we note that the integrand of the action 
\eqref{action0} vanishes for the solution \eqref{NES} and thus there is no contribution to the 
on shell action evaluated in the WDW patch from the bulk part. Therefore we are left with 
joint and boundary terms. If one
considers the corresponding WDW patch as that shown in the right panel of the figure 1
containing  two joint points at $\hat{r}_q$ and $\hat{r}_p$, one 
only has to compute  contributions of four joint points which results  in
the same form as that given in the equation \eqref{J0}. It is then easy to see that the resultant 
complexity approaches a constant at the late time contradicting the  result \eqref{LNE0}. 

A remedy to overcome this puzzle has been proposed in  
\cite{{Akhavan:2018wla},{Alishahiha:2018swh}} in which 
it was demonstrated that as soon as one sets a UV cutoff, it 
will automatically enforce us to have a cutoff behind the horizon. Moreover one also needs  
to consider the contribution of the  counter term requires to have finite free energy  evaluated 
on this cutoff. As a result  when 
we set the UV cutoff for the AdS$_2$  geometry at $r=r_{\rm Max}$,  there will be a cutoff 
behind the horizon given by  $\hat{r}_0\sim\frac{L^3}{\hat{r}_{\rm Max}^2}$. The presence 
of this cutoff removes the joint point $\hat{r}_q$ from the WDW patch leaving us 
with a space like boundary. Therefore we will have to compute contributions of five joint 
points together with  boundary terms evaluated on the cutoff surface $\hat{r}_0$.

Denoting the normal vectors to the null and space like boundaries by
 \be
 k_1=\alpha\left(\partial_{\hat{t}}+\frac{1}{\hat{f}(\hat{r})}\partial_{\hat{r}}\right),\;\;\;\;\;
 k_2=\beta\left(\partial_{\hat{t}}-\frac{1}{\hat{f}(\hat{r})}\partial_{\hat{r}}\right),\;\;\;\;\;
n_r=\frac{1}{\sqrt{\hat{f}(\hat{r}_0)}}\partial_{\hat{r}},
   \ee
 with $\hat{f}(r)=\left(\frac{\hat{r}^2}{L^2}-1\right)h(r_e)$, the joint contributions are 
\bea
I^{\rm joint }&=&\frac{V_{d-1}L^{d-1}}{8\pi G_N r_e^{d-1}}
 \left(\log\frac{\alpha\beta }{|f(\hat{r}_p)|}+
  \log\frac{\alpha}{\sqrt{|f(\hat{r}_0)|}}+
   \log\frac{\beta }{\sqrt{|f(\hat{r}_0)|}}-
  2 \log\frac{\alpha\beta }{|f({\hat{r}_{\rm Max})|}} \right)\cr &&\cr
&=&\frac{V_{d-1}L^{d-1}}{8\pi G_N r_e^{d-1}}
 \left(2 \log {|f({\hat{r}_{\rm Max})|}}-\log{|f(\hat{r}_p)|}-
  \log{{|f(\hat{r}_0)|}} \right)\,.
\eea
 Note  that
 the parameters $\alpha$ and $\beta$ drop from the final expression 
of the action of joint points and therefore we do not need extra boundary terms to remove
the ambiguity associated with the normalization of null vectors. 

As for the  boundary part, using the Affine parameterization for the null boundaries
their contributions vanish and we only need to consider boundary terms defined on the
 space like boundary behind the horizon. The corresponding 
 boundary terms consist of two terms: the standard Gibbons-Hawking term and the  
 counter term required to get  finite  free energy  for  the near horizon solution 
 \eqref{NES}\footnote{ The second term in this expression  is indeed the counter 
 term \eqref{CT1} written for the near extremal solution.}
\be\label{CT}
I^{{\rm surf}}=I^{\rm GH}+I^{\rm c.t.}=-\frac{1}{8\pi G_N}\int d^{d-1}x d\hat{t}\,\sqrt{ |h|} 
\left(K-\frac{\sqrt{h(r_e)}}{L}\right)\bigg|_{\hat{r}_0}\,,
\ee
that results in 
\be
I^{{\rm surf}}=\frac{V_{d-1}L^{d-1}h(r_e)}{8\pi G_Nr_e^{d-1}} \left(\frac{\hat{r}_0}{L^2}+\frac{1}
{L}\right) (\tau+2\hat{r}^*(\hat{r}_{\rm Max})-2\hat{r}^*(\hat{r}_0))\,.
\ee
Therefore altogether one arrives at
\be
I^{\rm total}=\frac{V_{d-1}L^{d-1}}{8\pi G_N r_e^{d-1}}
 \left(4 \log \frac{\hat{r}_{\rm Max}}{L}-\log (1-\frac{\hat{r}_p^2}{L^2}) \right)+
\frac{V_{d-1}L^{d-1}}{8\pi G_Nr_e^{d-1}}\frac{h(r_e)}
{L}(\tau+2\hat{r}^*(\hat{r}_{\rm Max})-2\hat{r}^*(\hat{r}_0))\,, 
  \ee
  whose time derivative is
 \be
\frac{dI^{\rm total}}{d\hat{\tau}}=\frac{V_{d-1}L^{d-2}}{8\pi G_N }\,
\frac{h(r_e)}{r_e^{d-1}} \left( 1+\frac{\hat{r}_p}{L} \right) ,
  \ee  
that reduces to  \eqref{LNE0} at the late time  when  $\hat{r}_p
\rightarrow L$.

To conclude we note that in order to get a consistent result for the rate of  complexity
growth for near horizon  solution \eqref{NES}, one needs to consider a cutoff behind the
horizon whose value is fixed by the UV cutoff and also the known boundary term (eq. \eqref{CT1})
 needed
to  make the free energy finite should be evaluated on this cutoff.


\section{Complexity of Charged Vaidya solution}

In this section we would like to compute complexity for a charged Vaidya solution\footnote{
Using CA proposal the complexity for neutral Vaidya metric
 has been already studied in several papers  including\cite{{Moosa:2017yvt},
 {Alishahiha:2018tep},{Chapman:2018dem},{Chapman:2018lsv},{Tanhayi:2018gcj},
 {Fan:2018xwf}} }.
This solution can be thought of as collapsing charged null shell that produces 
a charged black brane. The solution can be found from the action \eqref{action0}
by adding a proper extra charged  matter field. Indeed the Vaidya geometry we are looking 
for is sourced by an energy momentum tensor and a current density of a massless 
null charged matter. Therefore the equations of motion \eqref{EOM} should be 
modified as follows 
\bea
R_{\mu\nu}-F_{\alpha\mu}
F^{\alpha}_{\nu}-\frac{1}{2}\left(R+\frac{d(d-1)}{L^2}-\frac{1}{2}F^2\right)g_{\mu\nu}=
T_{\mu\nu}^{\rm ext},\;\;\;\;\partial_\alpha\left(\sqrt{-g}F^{\alpha\beta}\right)=(J^{\rm ext})^\beta.
\eea
In the present case the non-zero components of the energy momentum and charge current 
are $T_{vv}^{\rm ext}$ and $J_v^{\rm ext}$, that  assuming to have  time dependent mass and 
charge,  are given by  \cite{Galante:2012pv}
\be
T_{vv}^{\rm ext}=\frac{d-1}{2r}\frac{\partial f(r,v)}{\partial v},\;\;\;\;\;\;\;\;J_v^{\rm ext}
=\sqrt{(d-1)(d-2)}\,r^{d-1}\frac{\partial Q(v)}{\partial v}\,.
\ee
The  charged  Vaidya metric we are going to consider is\footnote{
Complexity for charged Vaidya metric has also been studied in \cite{Jiang:2018tlu}. 
Of course our study is different  than  this paper  where the author have considered a
charged black brane in the present of a null neutral collapsing matter that results
in another charged black brane with different mass.}
\be
ds^2=\frac{L^2}{r^2}\left(-f(r)dv^2-2 dr dv+d\vec{x}^2\right),\;\;\;\;\;\;A_v=\theta(v)
\sqrt{\frac{d-1}{d-2}}QL \;(r_+^{d-2}-r^{d-2}),
\ee
with
\be
f(r)=1+\theta(v)\left(-m r^d+{Q^2} r^{2(d-1)}\right),
\ee
where $\theta(v)$ is the step function. Therefore for $v<0$ one has an AdS solution while for $v>0$ it is the AdS-RN  black brane
we considered in the previous section.
The corresponding Penrose diagram is shown in  the 
figure 2 (see \cite{Callebaut:2014tva}).
\begin{figure}
\begin{center}
\includegraphics[scale=0.9]{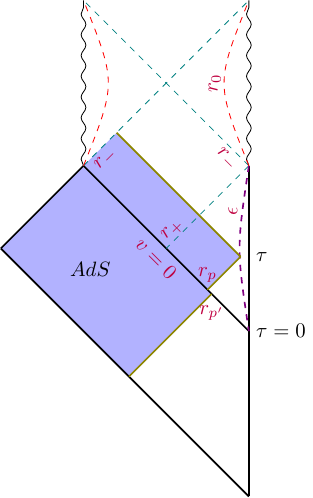}
\end{center}
\caption{ Penrose diagram for Vaidya AdS Reissner-Nordstr\"om
 spacetime. The figure is taken from \cite{Callebaut:2014tva}. The WDW patch is shown by
 blue color and the red dashed lines are 
 behind the horizon cutoff.  Note that due to the shock wave there is a displacement in the
 null boundary of WDW patch when crosses the shock wave. The displacement is shown 
 by joint points  $r_{p'}$ and $r_{p}$ in this figure.
 }
\label{fig:A}
\end{figure}

To compute the complexity for this model one needs to evaluate on shell action on the 
WDW patch depicted in the figure 2.\footnote{To draw the WDW patch in the present case 
we are motivated by the charged black hole where the corresponding WDW patch cannot
probe the region behind the inner horizon.  Indeed, it is clear that there is no way that 
the WDW patch can reach the region behind the inner horizon at the late time. 
This fact is also confirmed by 
the late time behavior of complexity growth.}
Following our notation the null boundaries of  the corresponding WDW patch are 
given by constant $v$ and $u$. To proceed, it is useful  
to decompose the WDW patch into three parts:  $v<0$ and $v>0$ and $v=0$ parts. 
It is known that the action on  null shell gives zero contribution
(see {\it e.g.} \cite{Chapman:2018dem})  and therefore we will 
 consider two parts given by $v>0$ and $v<0$.
The corresponding null boundaries of these two parts are 
\bea
v>0&:&\;\;\;\;\;\;\;\;\;\;v=0,\;\;\;\;v=\tau,\;\;\;\;\tau-v=2\int_\epsilon^{r(v,\tau)}\frac{dr}{f(r)},\cr &&\cr
v<0&:&\;\;\;\;\;\;\;\;\;\;v=0,\;\;\;\;v=-\infty,\;\;\;\; r(v,\tau)=r_{p'}(\tau)-\frac{1}{2}v\,,
\eea
where $\epsilon$ is the UV cutoff. 

Actually for $v<0$ the solution  is a pure AdS$_{d+1}$ solution and we will have to compute 
three different contributions to the on shell action that come from bulk, joint and ambiguity
terms of the action. Note that since all boundaries are null their contributions are zero
using the Affine parametrization to parametrize the null directions. It is then straightforward to 
compute the non-zero contributions. In particular for the bulk term one finds
\bea
I^{\rm bulk}_{v<0}=-\frac{dL^{d-1}}{8\pi G_N}\int d^{d-1}x\;
\int_{-\infty}^0 dv\; \int_{r_{p'}(\tau)-\frac{1}{2}v}^\infty \frac{dr}{r^{d+1}}
=-\frac{V_{d-1}L^{d-1}}{4\pi G_N}\frac{1}{d-1}\;\frac{1}{r^{d-1}_{p'}(\tau)}.
\eea
While using the null vectors \eqref{NV} the contribution of joint points is 
\be
I^{\rm joint }_{v<0}=-\frac{V_{d-1}}{8\pi G_N} 
\frac{L^{d-1}}{r_{p'}^{d-1}(\tau)}\log \frac{\alpha\beta r^2_{p'}(\tau)}{L^2}.
\ee
Finally one has a contribution from the ambiguity term that is 
\bea
I^{\rm amb}=\frac{V_{d-1}L^{d-1}}{8\pi G_N}\left(\frac{\log\frac{\alpha \beta r_{p'}^2(\tau)}{L^2}}
{r_{p'}^{d-1}(\tau)}
+\frac{2}{(d-1)r_{p'}^{d-1}(\tau)}\right).
\eea
Putting all terms together we find that on shell action vanishes 
for the  $v<0$ part of WDW patch. Therefore all nonzero contributions to the on 
shell action come from $v>0$ part.  Indeed in this part for the bulk term one has
\be
I^{\rm bulk}_{v>0}=\frac{L^{d-1}}{8\pi G_N}
\int d^{d-1}x\; \int_0^{\tau} 
dv\int_{r(v,\tau)}^{r_-}dr\, \left(\frac{-d}{r^{d+1}}+Q^2(d-2)r^{d-3}\right)=-
\frac{V_{d-1}L^{d-1}}{8\pi G_N}\int_0^{\tau}dv\,\frac{f(r(v,\tau))}
{r^{d}(v,\tau)}.
\ee 
On the other had using the fact that $\frac{dr(v,\tau)}{d\tau}=\frac{1}{2}f(r(v,\tau))$
one can perform the above integration resulting in
\be
I^{\rm bulk}_{v>0}=\frac{V_{d-1}L^{d-1}}{8\pi G_N}\left(\frac{2}{(d-1)r_p^{d-1}(\tau)}-
\frac{2}{(d-1)\epsilon^{d-1}}\right)\,.
\ee
Here we have used the fact that $r(\tau,\tau)=\epsilon, r(0,\tau)=r_p(\tau)$. It is important to note that
due to the null shock wave at $v=0$ there would be a displacement in the null boundary and 
the joint point $p$ is not necessary at the same point $p'$ we have considered in the AdS part
for $v<0$\cite{Chapman:2018dem}. Of course it does not affect our results since we 
compute the contribution of 
the two parts $v<0$ and $v>0$ separately.  

The $v>0$ part of the WDW has four null  boundaries whose contributions to 
the on shell action vanish using Affine parameterization for the null directions. Therefore 
the remaining part needs to be computed is the contribution of joint points. 

In this part of the WDW patch there are four joint points two of which are located 
at $r=r_-$, one at $r_p$ and one at $r=\epsilon$. Although the contributions of two later 
points can be easily computed using the coordinate systems we have been using so far, for
those at $r_-$ it finds useful to utilize the following coordinate system 
\be
U=-e^{-\frac{1}{2}f'(r_-)u},\;\;\;\;\;\;\;\;\;\;V=e^{\frac{1}{2}f'(r_-)v},
\ee 
by which the corresponding  two joint points are located at $(U,V)=(\zeta,V_0)$ and  $(U,V)=(\zeta,V_\tau)$, for
$\zeta\rightarrow 0$. Denoting the $r$ coordinate associated with these points by $r_0$ and
$r_\tau$, respectively, and using the normal vectors \eqref{NV} one gets the following expression 
for the contribution of  the joint points 
\bea
I^{\rm joint }_{v>0}=\frac{V_{d-1}L^{d-1}}{8\pi G_N} \left(\frac{\log\frac{\alpha\beta r^2_p(\tau)}{L^2f(r_p)}}{r_p^{d-1}(\tau)}-\frac{\log\frac{\alpha\beta \epsilon^2}{L^2f(\epsilon)}}{\epsilon^{d-1}}
+\frac{\log\frac{\alpha\beta r^2_\tau}{L^2f(r_\tau)}}{r_\tau^{d-1}}-\frac{\log\frac{\alpha\beta r^2_0}{L^2f(r_0)}}{r_0^{d-1}}\,
\right).
\eea
On the other hand using the fact that \cite{Agon:2018zso}\footnote{Here  $c_0=\psi^{(0)}(1)-\psi^{(0)}(\frac{1}{d+1})$ is a 
positive number and $\psi^{(0)}(x)=\frac{\Gamma'(x)}{\Gamma(x)}$ is the 
digamma function.}
\be
\log f(r)=\log |UV|+c_0 +{\cal O}(UV),\;\;\;\;\;{\rm for}\;\;UV\rightarrow 0,
\ee
the above expression reads
\bea
I^{\rm joint }_{v>0}=\frac{V_{d-1}L^{d-1}}{8\pi G_N} \left(\frac{\log\frac{\alpha\beta r^2_p(\tau)}{L^2}}{r_p^{d-1}(\tau)}-\frac{\log\frac{\alpha\beta \epsilon^2}{L^2}}{\epsilon^{d-1}}
-\frac{\log f(r_p)}{r_p^{d-1}(\tau)}+\frac{f'(r_-)\tau}{2r_-^{d-1}}
\right).
\eea
Here we have also used the fact that $\{r_0,r_\tau\}\approx r_-$. 
Finally the  contribution of ambiguity term is
\bea
I^{\rm amb}_{v>0}&=&\frac{V_{d-1}L^{d-1}}{8\pi G_N}\left(-\frac{\log\frac{\alpha r_-}{L}}{r_-^{d-1}}
-\frac{1}{(d-1)r_-^{d-1}}+\frac{\log\frac{\alpha \epsilon}{L}}{\epsilon^{d-1}}
+\frac{1}{(d-1)\epsilon^{d-1}}\right)\cr &&\cr
&&+\frac{V_{d-1}L^{d-1}}{8\pi G_N}\left(-\frac{\log\frac{\beta r_p(\tau)}{L}}{r_p^{d-1}(\tau)}
-\frac{1}{(d-1)r_p^{d-1}(\tau)}+\frac{\log\frac{\beta \epsilon}{L}}{\epsilon^{d-1}}
+\frac{1}{(d-1)\epsilon^{d-1}}\right)
\cr &&\cr
&&+\frac{V_{d-1}L^{d-1}}{8\pi G_N}\left(-\frac{\log\frac{\alpha r_p(\tau)}{L}}{r_p^{d-1}(\tau)}
-\frac{1}{(d-1)r_p^{d-1}(\tau)}+\frac{\log\frac{\alpha r_-}{L}}{r_-^{d-1}}
+\frac{1}{(d-1)r_-^{d-1}}\right)\cr &&\cr
&=&\frac{V_{d-1}L^{d-1}}{8\pi G_N}\left(-\frac{\log\frac{\alpha \beta r_p^2(\tau)}{L^2}}
{r_p^{d-1}(\tau)}
-\frac{2}{(d-1)r_p^{d-1}(\tau)}+\frac{\log\frac{\alpha\beta \epsilon^2}{L^2}}{\epsilon^{d-1}}
+\frac{2}{(d-1)\epsilon^{d-1}}\right).
\eea

As we already mentioned the  contribution of $v<0$ part vanishes and therefore 
the total  on shell is
\be
I^{\rm total}=I^{\rm bulk}_{v>0}+I^{\rm joint}_{v>0}+I^{\rm amb}_{v>0}=\frac{V_{d-1}L^{d-1}}{8\pi G_N}\bigg(\frac{f'(r_-)}{2r_-^{d-1}}\tau-\frac{\log f(r_p(\tau))}{r_p^{d-1}(\tau)}\bigg).
\ee
It is then easy to compute the time derivative of on shell action 
\be
\frac{dI^{\rm total}}{d\tau}=\frac{V_{d-1}L^{d-1}}{16\pi G_N}\left(\frac{f'(r_-)}{r_-^{d-1}}-
\frac{f'(r_p(\tau))}{r_p^{d-1}(\tau)}+\frac{d-1}{ r_p^d(\tau)} f(r_p(\tau)) \log |f(r_p(\tau))|\right) .
\ee
Note that at late time where $r_p(\infty)=r_+$ one gets
\be
\left(\frac{dI^{\rm total}}{d\tau}\right)_\infty=\frac{V_{d-1}L^{d-1}}{16\pi G_N}\left(\frac{f'(r_-)}{r_-^{d-1}}-
\frac{f'(r_+)}{r_+^{d-1}}\right) ,
\ee
that is the same as that of eternal black brane given in the equation \eqref{R1}. It is
worth mentioning that since $r_p\leq r_+$ the logarithmic term is always negative leading 
to the fact that the complexity growth approaches its late time value from below respecting 
the Lloyd's bound. The full time dependence of complexity growth is depicted in the figure 3.  
\begin{figure}
\begin{center}
\includegraphics[scale=0.7]{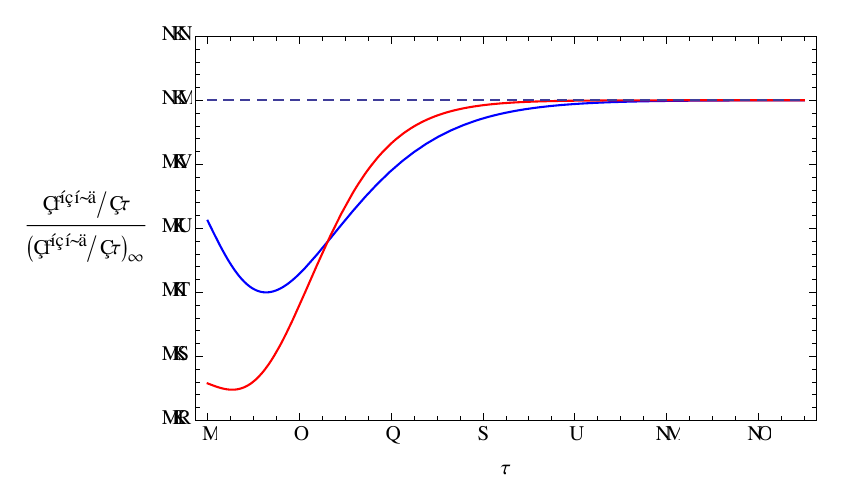}
\end{center}
\caption{Rate of the complexity growth for charged Vaidya geometry with
$m=1,Q=\frac{1}{3}$ (blue) and $m=1,Q=\frac{1}{10}$ (red). The  growth respects  
the Lloyd's bound as in the uncharged case. }
\end{figure}

Since all non-zero contributions to the on shell action come from the $v>0$ part where 
the geometry is essentially an AdS-RN black brane, one may also study  a case where 
the corresponding black brane is in the stage of near extremal. 
Of course in order to get
such a geometry the charge and the energy of infalling  null matter should be fine tuned.
Since in the present case the late time behavior of complexity is the same as that of 
two-sided black brane
considered in the previous section, taking the corresponding near horizon limit 
would lead to the same expression given in the equation \eqref{LNE0} too.

One could also try to directly compute the complexity from the near horizon solution.
We note, however, that in general it is not straightforward to write a solution that is 
a $d+1$ dimensional AdS for $v<0$ and an AdS$_2 \times \mathbb{R}^{d-1}$ for 
$v>0$. Nonetheless as far as the computation of complexity is concerned it is possible to 
proceed to find the complexity for near horizon solution. 
Actually as we have already seen the contribution to 
the complexity from the $v<0$ part where we have an AdS$_{d-1}$ is zero and therefore 
we are left with just  $v>0$ part. On the other hand in this part the geometry is
the  near horizon limit of near extremal case which we have considered in the previous section.
Therefore we will also get the same expression for the late time behavior of  complexity growth
as  \eqref{LNE0}. 

It is worth noting that since the near horizon limit of the next extremal geometry results in a
geometry containing an AdS$_2$ geometry, in order to compute the corresponding complexity 
the cutoff behind the horizon is required to get the correct result.

Another interesting case related to what we have considered here is  to study complexity 
for a process of  creating  a near extremal RN black hole from the extremal one by adding 
an in falling shock wave. Such a geometry has been studied in \cite{Fabbri:2000xh}.

\section{Conclusions}

In this paper we have studied complexity for near horizon limit of near extremal charged 
black brane solutions. The aim was to compute the corresponding complexity from two 
different approaches. In the first approach we have taken the  limit from the complexity of 
a charged black brane while in the second approach we have computed complexity for a 
near horizon solution.

We have observed that in order to get the same result from both approaches one has to 
consider a cutoff  behind the horizon  whose value  is given by  the UV cutoff. 
Moreover it is crucial to add the contribution of the  counter term required to have 
finite free energy evaluated on the cutoff surface behind the horizon.

It is also important  to note that in order to find meaningful results one has to carefully take 
the near horizon limit of the near extremal black brane. Doing so, the resultant metric  has 
an AdS$_2$ factor in the Rindler coordinates.  It is worth noting  that although we have 
computed the complexity from $d+1$ dimensional theory point of view, 
one could have reduced the near horizon solution into two
dimensions where we would have gotten the  generalized Jackiw-Teitelboim gravity 
(see for example \cite{Navarro:1999qy}).  The corresponding two dimensional gravity may be obtained by a dimensional reduction using the following general ansatz 
\be
ds^2=d\hat{s}_2^2+\psi^2\ d\vec{x}^2.
\ee
Plugging this metric in to the action \eqref{action0} one gets (see {\it e.g.} \cite{Navarro:1999qy}
for more details)\footnote{ It is important to note that in what follows we only consider the case
where the higher dimensional theory has an AdS$_2$ factor and the reduction is along other 
directions and therefore we will end up with a generalized Jackiw-Teitelbiom gravity whose 
vacuum solution is AdS$_2$ geometry. This is indeed the case where we have problem with 
complexity computations. In this paper we proposed that the non-zero complexity can be 
achieved by making use of the behind horizon cutoff, while the authors of 
\cite{{Brown:2018bms},{Goto:2019}} have proposed another
approach. Interestingly enough in both cases the main role is played by a certain boundary 
term which could eventually be related to each other. It would be interesting to explore this
relation. }
\be
I=\frac{V_{d-1}}{16\pi  G_N}\int d^2x\sqrt{-\hat{g}_2}\psi^{d-1}\left(\hat{ R}_2+(d-1)(d-2)|
\hat{\nabla} \psi|^2\psi^{-2}+\frac{d(d-1)}{L^2}-\frac{1}{2} \hat{F}_2^2\right)\,.
\ee

Then the complexity could have been computed 
from two dimensional point of view. We note that complexity for the (generalized) 
Jackiw-Teitelboim gravity has been studied 
in \cite{{Brown:2018bms},{Alishahiha:2018swh},{Goto:2019}}.  In particular the authors of 
\cite{{Brown:2018bms},{Goto:2019}} have studied four dimensional charged black hole 
reduced into 
two dimensions. It was shown that in order to find the desired late times linear growth for 
complexity the contribution of certain boundary term must be taken into account.
The corresponding boundary term is  ( here we have  written the corresponding term 
for arbitrary dimensions) 
\be\label{BT}
I=\frac{1}{8\pi G_N}\int d^{d}x\,\sqrt{|h|}\,\,n_\mu F^{\mu\nu} A_\nu\,,
\ee
where $n_\mu$ is normal vector to the boundary. Actually this boundary term is the one needed 
to impose  boundary condition on the gauge field rather than fixing the 
potential at the boundary.  Using the equation of motion the above term may 
be recast to a volume term to be evaluated on the WDW patch. 
Interestingly enough the final result for  complexity growth coincides with what we 
have obtained using behind the horizon cutoff. Actually it is straightforward to see that the 
boundary term \eqref{BT} evaluated on the behind the horizon cutoff results in
\be
I=-\frac{1}{8\pi G_N}\int d^{d}x\,\sqrt{|h|}\,\frac{\sqrt{h(r_e)}}{L}\,,
\ee
that is exactly the counter terms we have considered. Of course in our picture our motivation to 
consider this counter term was given by the holographic renormalization 
 It would be interesting to understand the relation  
between these two approaches better. 

As a final remark we would like to make a comment on the behind the horizon cutoff for the 
charged black holes. Although in this case we have not obtained the relation between UV 
cutoff and the behind horizon cutoff, since for small UV cutoff we would expect that the
corresponding cutoff tends to infinity where the singularity is located the behind horizon 
cutoff should be behind the inner horizon, as shown by red dashed lines in the left panel 
of figure 1 and figure 2. Therefore it does not have any direct effect in the WDW patch and 
thus  complexity.

Indeed it is consistent with the result of \cite{Hartman:2018tkw} where a charged black hole 
at a finite cutoff has been studied. In this case one could see that unlike  the neutral
black holes the complexity is not affected by the cutoff. 

On the other hand going into the near horizon the region in which the behind horizon cutoff is 
located will be excluded from the spacetime and we will have to reconsider the cutoff 
in the new coordinates denoted by $(\hat{t},\hat{r})$  in present paper.  More precisely in this 
case the obtained spacetime has  an AdS$_2$ factor  and the corresponding cutoff is depicted 
by dashed red line in the right panel of figure 1.

\subsection*{Acknowledgments}

The authors would like to kindly thank A. Akhavan, M. H. Halataei,  
M. R. Mohammadi Mozaffar,  A. Naseh,  F. Omidi, and  M.H. Vahidinia for useful comments  and 
discussions  on related topics.

\end{document}